\documentclass[fleqn,twoside]{article}
\usepackage{espcrc2}

\usepackage{float}

\usepackage{graphicx}
\usepackage[figuresright]{rotating}

\newcommand{\AmS}{{\protect\the\textfont2
  A\kern-.1667em\lower.5ex\hbox{M}\kern-.125emS}}

\def\cO{  {\cal O}  }

\title{
Analytic results for multiloop scattering amplitudes
\thanks{Talk given at 
{\it Loops and Legs in Quantum Field Theory}, DESY, 2012;
{\it Scattering Amplitudes: from QCD to maximally supersymmetric Yang-Mills theory and back}, ECT, 2012;
{\it Calculations for Modern and Future Colliders}, CALC, 2012;
{\it The Geometry of Scattering Amplitudes, BIRS}, 2012.
}
}

\author{J. M. Henn
\address[MCSD]{Institute for Advanced Study, 
       Einstein Drive, Princeton, NJ 08540, 
        USA \\
      }%          
                }

\begin{document}

\begin{abstract}
The velocity-dependent cusp anomalous dimension is by definition the
ultraviolet (UV) anomalous dimension of a Wilson loop with a cusp.
It appears in many physically interesting processes. 
In this talk we present recent progress in the  analytic calculation of
that quantity.
%We present new perturbative results obtained from a relation to the Regge 
%limit of massive amplitudes. 
%We propose a new scaling limit and resume
%the perturbative series in this case, finding agreement with a string theory
%calculation.
%Finally, we obtain a exact result for small angles.

\vspace{1pc}
\end{abstract}

\maketitle

\section{INTRODUCTION}

Wilson loops are very fundamental and important quantities in gauge theories \cite{PolyakovCusp}. 
They are essential for defining nonlocal gauge invariant quantities.
They also contain information about local operators via the operator product expansion.
Moreover, it turns out that certain Wilson loops defined for
specific contours appear as an effective description of certain physical
processes. 

The velocity-dependent cusp anomalous dimension is a case in point.
It is defined as the ultraviolet (UV) anomalous dimension of a Wilson loop with a cusp.
It appears in many physically interesting situations. 
For example, in heavy quark effective theory (HQET), it describes the infrared (IR) 
divergences of massive form factors and scattering amplitudes, see e.g. 
\cite{Korchemsky:1991zp,Neubert:1993mb,Manohar:2000dt,Grozin:2004yc}. 

\begin{figure}
\centering\includegraphics[scale=0.6]{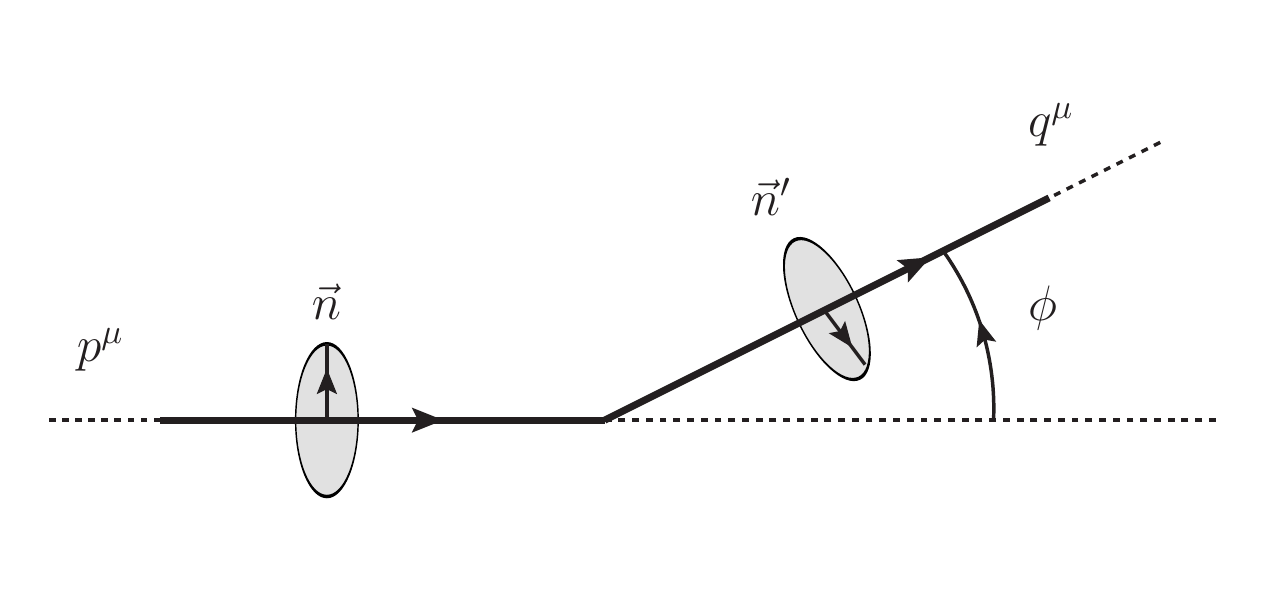}
\vspace{-1.5cm}
\caption{A Wilson line that makes a turn by an angle $\phi$ in Euclidean space.
The two segments are directed along $p^{\mu}$ and $q^{\mu}$, respectively.
The vectors $\vec{n}$ and $\vec{n}'$ are internal vectors that determine the
coupling to the scalars $\vec{\Phi}$, see eq. (\ref{susy-loop}).}
\label{fig:WL}
\end{figure}

In a $\mathcal{N}=4$ super Yang-Mills (SYM), it is natural to define the locally supersymmetric
Wilson loop operator \cite{Maldacena:1998im,Rey:1998ik}
\begin{eqnarray}\label{susy-loop}
W \sim {\rm Tr}\left[P \exp \left({ i \oint  A^{\mu} \dot{x}_{\mu} + \oint |dx| \vec{n} \cdot \vec{\Phi}} \right) \right]\,,
\end{eqnarray}
where $\vec{n}$ is a vector on $S^{5}$. It parametrizes the 
coupling of the Wilson loop to the six scalars $\vec{\Phi}$ of the theory.
We consider as the integration contour a cusp formed by two segments
along directions (momenta) $p^{\mu}$ and $q^{\mu}$,
and allow the two segments to couple to the scalars through vectors $\vec{n}$ and $\vec{n}'$, see Fig.~\ref{fig:WL}.
Then, the vacuum expectation value $\langle W \rangle$ of the Wilson loop will depend on the angles
\begin{eqnarray}
\cos \phi  = \frac{ p \cdot q}{\sqrt{p^2 q^2}} \,,\quad  \cos \theta = \vec{n} \cdot \vec{n}' \,,
\end{eqnarray}
as well as on the `t Hooft coupling $\lambda=g^2 N$, and the rank of the gauge group $N$.

If $\Lambda_{\rm UV}$ and $\Lambda_{\rm IR}$ are short and large distance cutoffs, respectively,
then the divergent part of the vacuum expectation value of the Wilson loop takes the form \cite{PolyakovCusp,Brandt:1981kf}
\begin{eqnarray}
\left\langle W \right\rangle \sim \exp{ \left[ - \log \frac{\Lambda_{\rm UV }}{\Lambda_{\rm IR}} \, \Gamma_{\rm cusp} + \ldots \right] } \,.
\end{eqnarray}
This defines the cusp anomalous dimension $\Gamma_{\rm cusp}(\phi, \theta, \lambda, N)$.

%\section{OUTLINE}

We will begin by describing the dependence of $\Gamma_{\rm cusp}$ on the angles
$\phi$ and $\theta$ in section \ref{sec:kinematics}, and then introduce, in section \ref{sec:functions},  the set of
integral functions needed to express loop-level results.
We will then review perturbative results to two loops, and present the new three-loop result.
We present the light-like limit as a check in section \ref{sec:limit}, and describe our
method of calculation in \ref{methodcalculation}.
We then present an exact result for $\Gamma_{\rm cusp}$ at small angles in section \ref{sec:bremsstrahlung}.
In section \ref{sec:scaling}, we discuss a new scaling limit. 
Finally, we conclude and present future directions in section \ref{sec:conclusion}.

\section{KINEMATICS AND LIMITS}
\label{sec:kinematics}

For reasons that will become apparent presently, we will mostly be interested in the 
$\phi$ dependence of the Wilson loop.
It is convenient to introduce a new variable $x=e^{i \phi}$.
The computation we are considering is invariant under $\phi \to -\phi$. 
This corresponds to an inversion symmetry in $x$.

Note that the dependence of $\Gamma_{\rm cusp}$ on $\theta$ is simple. 
It can only occur through Wick contractions of scalars, and because of SO(6) invariance
it appears only through $\vec{n} \cdot \vec{n}' = \cos \theta$. 
Therefore, at $L$ loops, $\Gamma_{\rm cusp}$ is a polynomial in $\cos \theta$,
of maximal degree $L$. 
Having made this observation, we find that it convenient
to introduce the variable $\xi = (\cos \theta - \cos \phi)/(i \sin \phi)$. 
Two important cases are $\theta = 0$ (constant coupling to scalars) 
and $\theta=\pi/2$ (scalars on opposite edges are orthogonal to each other), 
which lead to $\xi = (1-x)/(1+x)$ and $\xi = (1+x^2)/(1-x^2)$, respectively.

There are several special cases of the angles that are of particular interest.
When the geometric angle $\phi$ and internal angle $\theta$ satisfy $\phi = \pm \theta$,
the anomalous dimension vanishes. Note that $\xi$ vanishes in this case.
For $\theta =0$, this corresponds to 
$\phi \to 0$, i.e. $x \to 1$, the case of a straight line.
The small angle limit is related to energy loss of an accelerated quark, and
is known exactly \cite{Correa:2012at,Fiol:2012sg}.
The limit $\phi \to \pi$, i.e. $x \to -1$, is related to the quark-antiquark potential.
This limit is subtle and requires a resummation of certain diagrams, see 
\cite{Erickson:1999qv,Pineda:2007kz,Correa:2012nk,Stahlhofen:2012zx}.

The above limits can be defined in Euclidean space.
There is an intrinsically Minkowskian limit that is also of interest. When
$\phi \to - i \varphi$, $\varphi \gg 1$, i.e. $x \to 0$, the cusp anomalous dimension diverges linearly
in the $\varphi$, to all orders in the coupling constant \cite{Korchemsky:1987wg}. The coefficient of the
linear divergence is the well-studied light-light cusp anomalous dimension; the latter can
also be obtained from the anomalous dimension of high spin operators  \cite{Korchemsky:1988si,Korchemsky:1992xv,Alday:2007mf}. 
The Wilson loop approach considered here is a very efficient way of computing this quantity.

\section{INTEGRAL FUNCTIONS}
\label{sec:functions}

\begin{figure}
\centering\includegraphics[scale=0.5]{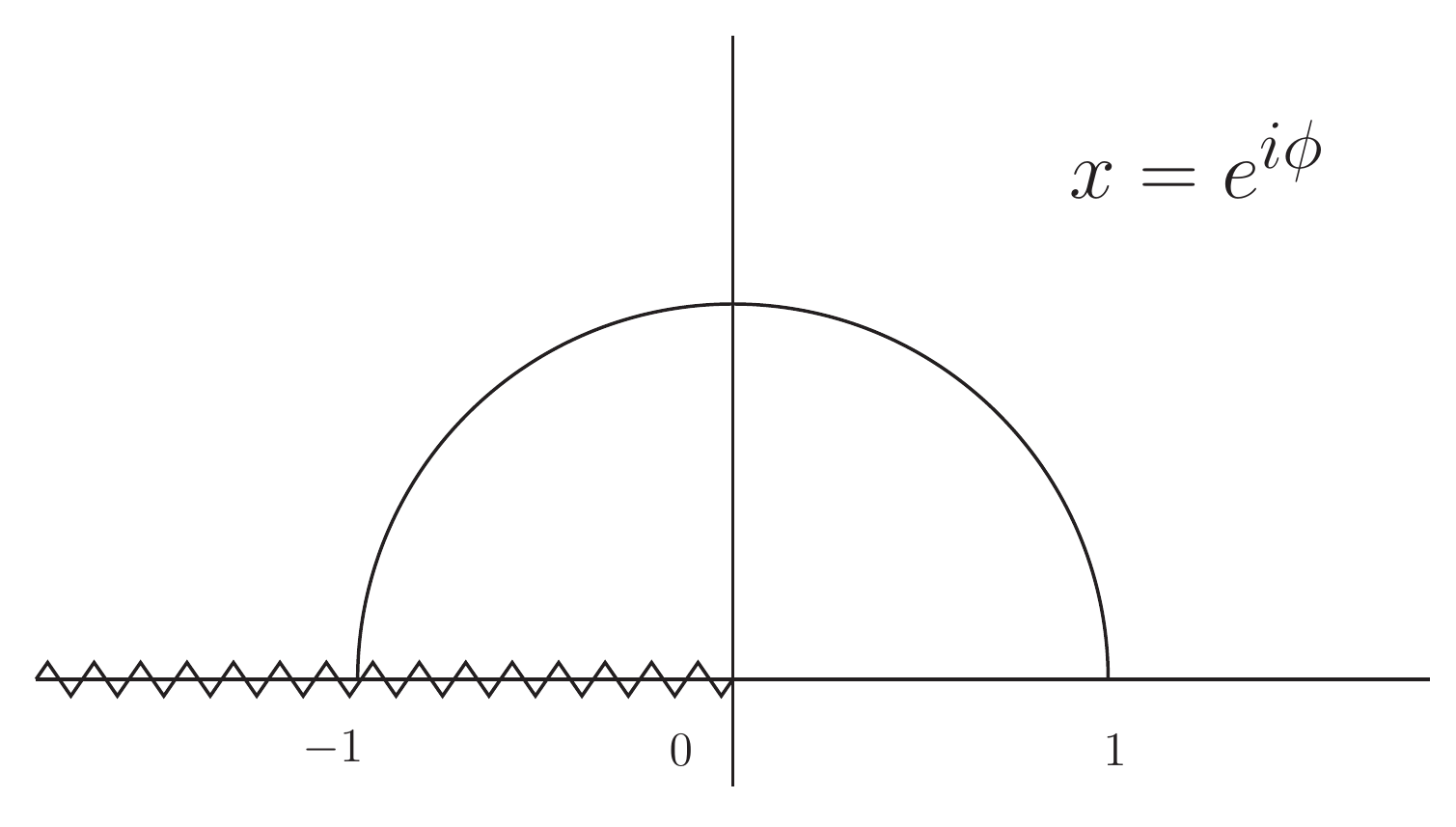}
\vspace{-1.0 cm}
\caption{$\Gamma_{\rm cusp}$ is real for $0<x<1$, and below threshold, where $x=e^{i \phi}$ is a phase.
Above threshold, we have $-1<x<0$, with $x$ having an infinitesimal imaginary part. The zigzag line denotes
a branch cut along the negative real axis. Recall that $\Gamma_{\rm cusp}$ has is an inversion symmetry $x\to 1/x$.}
\label{fig:xcomplex}
\end{figure}

Let us discuss the different kinematical regions for complex $x$.
It is useful to recall the relationship of $\Gamma_{\rm cusp}$ to IR
divergences of massive form factors, such as $\gamma^{*} \rightarrow e^+ e^-$, which
have the same analytical structure.

$\Gamma_{\rm cusp}$ is real in the Euclidean region $0<x<1$, and in the
physical region below threshold (of creating two massive particles), which is the unit circle (i.e. for $\phi$ real). The threshold is at
$x=-1$, and the physical region above threshold goes from $-1$ to $0$ along the negative real axis. 
See Fig.~\ref{fig:xcomplex}.
We find that all results can be expressed in terms of harmonic polylogarithms (HPL) \cite{Remiddi:1999ew}.
These functions are generalizations of classical polylogarithms, and appear naturally
in this problem. 
They are defined iteratively by
\begin{eqnarray}\label{defHPL}
H_{a_1 , a_2 , \ldots, a_n }(x) = \int_0^x f_{a_1}(t) H_{a_2, \ldots, a_n }(t) \, dt \,,
\end{eqnarray}
where the integration kernels are defined as
${f_{1}(x) = (1-x)^{-1}}$,  ${f_{0}(x) = x^{-1}}$, and ${f_{-1}(x) = (1+x)^{-1}}$.
The degree (or weight) one functions needed to start the recursion are defined as
\begin{eqnarray}
H_{1}(x) &=& - \log(1-x) \,,\qquad H_{0}(x) = \log(x) \,, \nonumber \\ && H_{-1}(x) = \log(1+x) \,.
\end{eqnarray}
The subscript of $H$ is called the weight vector. A common abbreviation is to replace occurrences
of $m$ zeros to the left of $\pm1$ by $\pm(m+1)$. For example, $H_{0,0,1,0,-1}(x) = H_{3,-2}(x)$.

HPLs have simple properties under certain argument transformations, and one can use
their algebraic properties in order to make their asymptotic behavior manifest.
We refer the interested reader to ref. \cite{Remiddi:1999ew}. 
A very useful computer algebraic implementation has been given in ref. \cite{Maitre:2005uu}.

The perturbative results given in section \ref{sec:results} can be straight forwardly evaluated
numerically in the region $0<x<1$. Other regions can be reached by analytical continuation,
respecting the above branch cut properties. A related discussion is given in ref. \cite{Anastasiou:2006hc}
for the process $g  g \rightarrow h$ via a massive quark loop.

\section{RESULTS UP TO THREE LOOPS}
\label{sec:results}

The cusp anomalous dimension of the standard bosonic Wilson loop operator
was computed in QCD at two loops in the pioneering paper \cite{Korchemsky:1987wg}.
This result was later simplified \cite{Kidonakis:2009ev}, and recomputed in ${\mathcal N}=4$ SYM
for the supersymmetric loop operator defined in eq. (\ref{susy-loop})
in refs. \cite{Makeenko:2006ds,Drukker:2011za}.
The answer in ${\mathcal N}=4$ SYM can be written as an expansion in the `t Hooft coupling $\lambda$,
\begin{eqnarray}
\Gamma_{\rm cusp} = \sum_{L\ge 1} \left( \frac{ \lambda}{8 \pi^2 }\right)^L \, \Gamma_{\rm cusp}^{(L)} \,.
\end{eqnarray}
To two loops it is given by,
\begin{eqnarray}
\Gamma^{(1)}_{\rm cusp} &=& - \xi  \log x \,, \label{result1} \\
\Gamma^{(2)}_{\rm cusp} &=&  \frac{1}{3} \xi \left[  \log x \left( \log^2 x + \pi^2 \right) \right]  \nonumber \\
&& + \frac{1}{4} \xi^2 \left[ H_{1,1,1} + 2 H_{1,2} \right]\,.\label{result2}
\end{eqnarray}
Here and in the following, the HPLs are understood to have argument $1-x^2$.
In ref. \cite{Correa:2012nk}, we computed the three-loop value $\Gamma_{\rm cusp}^{(3)}$. 
Before outlining the method of calculation in section \ref{methodcalculation}, let us 
present the result.
Here we write it in
a very compact form, due to \cite{Henn:2012qz},
\begin{eqnarray}
\Gamma^{(3)}_{\rm cusp} &=& - \frac{1}{6}  \xi \log x \left( \log^2 x + \pi^2 \right)^2 \nonumber \\
&& \hspace{-1.5cm} -\frac{1}{2} \xi^2 [ 3 \zeta_3 H_{1,1} + \zeta_2 (2 H_{1,2}+4 H_{2,1}+3 H_{1,1,1}) \nonumber \\ 
&&\hspace{-1.5cm} + 2 H_{1,1,1,2}+\frac{3}{2} H_{1,2,1,1}+2 H_{2,1,1,1}+\frac{11}{4} H_{1,1,1,1,1} ] \nonumber \\
&&\hspace{-1.5cm} +\frac{1}{4} \xi^3 [ 4 H_{1, 1, 3} + 4 H_{1, 2, 2} + 4 H_{1, 1, 1, 2} \nonumber \\
&&\hspace{-1cm} + 
 2 H_{1, 1, 2, 1} +2 H_{1, 2, 1, 1} + 
 3 H_{1, 1, 1, 1, 1}   ]\,. \label{result3}
\end{eqnarray}
Let us now discuss these formulas.

It is not hard to verify the expected inversion symmetry $x\to 1/x$, i.e. $\phi\to -\phi$,
using standard relations between HPLs of related arguments \cite{Remiddi:1999ew}. Note
that $\xi$ is antisymmetric under this transformation.

We see that at each loop order, we have functions of uniform degree of 
transcendentality (i.e. weight of harmonic polylogarithms) ${(2L-1)}$.

It is remarkable that the result can be written using argument $1-x^2$ (or $x^2$) only.
This is not the case in general for individual Feynman integrals. 
Moreover, we see that there are no relative signs within the coefficients
of the different powers of $\xi$ in the expressions above.
This property also holds at least up to $L=6$ for the  $\xi^L$ terms at $L$ loops,
which have recently been computed analytically \cite{Henn:2012qz}.

\section{RELATION TO LIGHT-LIKE CUSP ANOMALOUS DIMENSION}
\label{sec:limit}

Let us use the above three-loop result to analytically verify
the relation 
\begin{eqnarray}
\mathop{\lim}_{x  \to 0} \Gamma_{\rm cusp} = - \frac{1}{2} \log x \, \Gamma^{\infty}_{\rm cusp} + \cO(x^0) \,,
\end{eqnarray}
where $\Gamma^{\infty}_{\rm cusp}$ is the anomalous dimension of a light-like cusped Wilson loop \cite{Korchemskaya:1992je}.
The asymptotic limit $x \to 0$ is easy to perform on our results, since the logarithmic behavior of HPLs can always
be made manifest \cite{Remiddi:1999ew}.
From eqs. (\ref{result1}), (\ref{result2}) and (\ref{result3}) it is easy to verify its correct three-loop value \cite{Bern:2005iz}
\begin{eqnarray}
\Gamma^{\infty}_{\rm cusp} &=& 
2 \left(\frac{\lambda}{8 \pi^2} \right)
-2 \zeta_2 \left(\frac{\lambda}{8 \pi^2} \right)^2
 \nonumber \\
&& 
+11 \zeta_4 \left(\frac{\lambda}{8 \pi^2} \right)^3 + \cO(\lambda^4) \,.
\end{eqnarray}

\section{RELATION TO MASSIVE SCATTERING AMPLITUDES IN N=4 SYM}
\label{methodcalculation}

The velocity-dependent cusp anomalous dimension governs the infrared
divergences of massive scattering amplitudes. 
In $\mathcal{N}=4$ super Yang-Mills, one can introduce masses by giving a
vacuum expectation value to some of the scalar particles. 
In this way, at the planar level, one can define finite four-dimensional 
scattering amplitudes which have an exact dual conformal symmetry \cite{Alday:2009zm}.
The simplest case is the four-particle scattering amplitude $M_{4}(u,v)$, which is a function 
of two conformally invariant variables $u$ and $v$, which in turn are defined in terms of
the momenta and masses.

It was argued in \cite{Henn:2010bk,Henn:2010kb,Correa:2012nk} that the limit
$u \ll 1$ of this function is determined by the cusp anomalous dimension in the following way,
\begin{eqnarray}
\mathop{\lim}_{u  \to 0} \log M_{4}(u,v)  = \log u \, \Gamma_{\rm cusp}(v) + \cO(u^0) \,.
\end{eqnarray}
The relation between $v$ and the previously used variables is $v=x/(1-x)^2$.

Let us explain how we used this limit to compute $\Gamma_{\rm cusp}$ at
three loops \cite{Correa:2012nk}, starting from a known representation of the three-loop scattering
amplitude \cite{Bern:2005iz,Henn:2010bk}.
We observed that all known form factor and Wilson line integrals of this type
could be expressed in terms of harmonic polylogarithms of argument $x$.
Assuming that this holds for $\Gamma_{\rm cusp}$ at three loops reduced the calculation 
to the problem of determining a number of coefficients. We used Mellin-Barnes
techniques \cite{Henn:2010bk} to compute the asymptotic limit $x \to 0$, keeping
not only powers of $\log x$, but also terms suppressed by powers of $x$.
In this way, we were able to fix our ansatz completely.

We note that recently, our assumption was proven for two infinite classes
of integrals, and moreover that it is possible to directly evaluate
the corresponding integrals \cite{Henn:2012qz}.

\begin{figure}
\centering\includegraphics[scale=0.4]{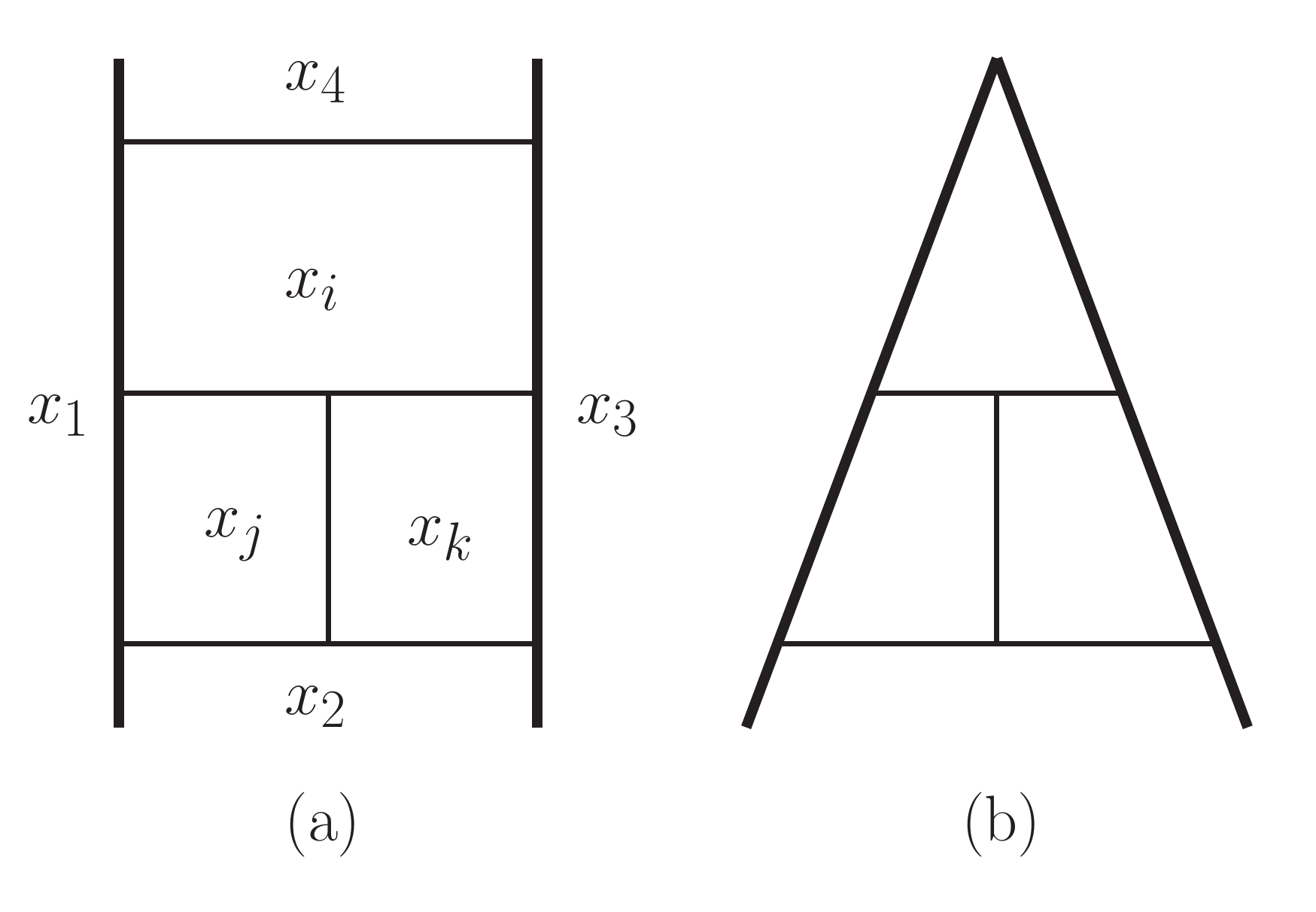}
\vspace{-1.0cm}
\caption{The form factor (or Wilson loop) integrand is obtained by taking the limit
where all dual integration points, $x_{i}, x_{j}, x_{k}$ in (a), come close to the external 
point $x_{2}$. Here we show a three-loop integral for the four-particle amplitude (a)
giving rise to contribution for the Wilson loop, shown in (b). Thick lines represent massive (or
eikonal) propagators, and thin lines represent massless propagators.}
\label{fig:example_integrand}
\end{figure}

We wish to emphasize that the relation to the four-particle amplitude discussed
above  can also be used at the
level of the integrand, where the dominant region of integration as $u \to 0$ 
corresponds to all dual loop momenta\footnote{Here we are using the
notion of dual (or region) variables for a planar graph.} 
approaching one of the external dual variables, say $x^{\mu}_{2}$.
In that region, one can simplify the integrand (to logarithmic accuracy)
to obtain integrals of form factor type. In the same spirit, one can
further approximate them in order to get a Wilson line representation.
We illustrate this simplification in Fig.~\ref{fig:example_integrand}.

In summary, this means that the integrand of the Wilson line calculation can
straightforwardly be obtained at higher loop orders
from the corresponding integrand of massless  scattering amplitudes 
\cite{ArkaniHamed:2010kv,Bourjaily:2011hi,Eden:2012tu},
extended properly to the massive case \cite{Alday:2009zm,Henn:2010bk}.
This should represent a useful starting point for calculations at higher 
loop orders, as well as for finding and proving conjectures about
the structure of $\Gamma_{\rm cusp}$, see e.g. section \ref{sec:scaling}.

\section{EXACT RESULT FOR SMALL \\ANGLES}
\label{sec:bremsstrahlung}

\begin{figure}[t]
\centering\includegraphics[scale=0.8]{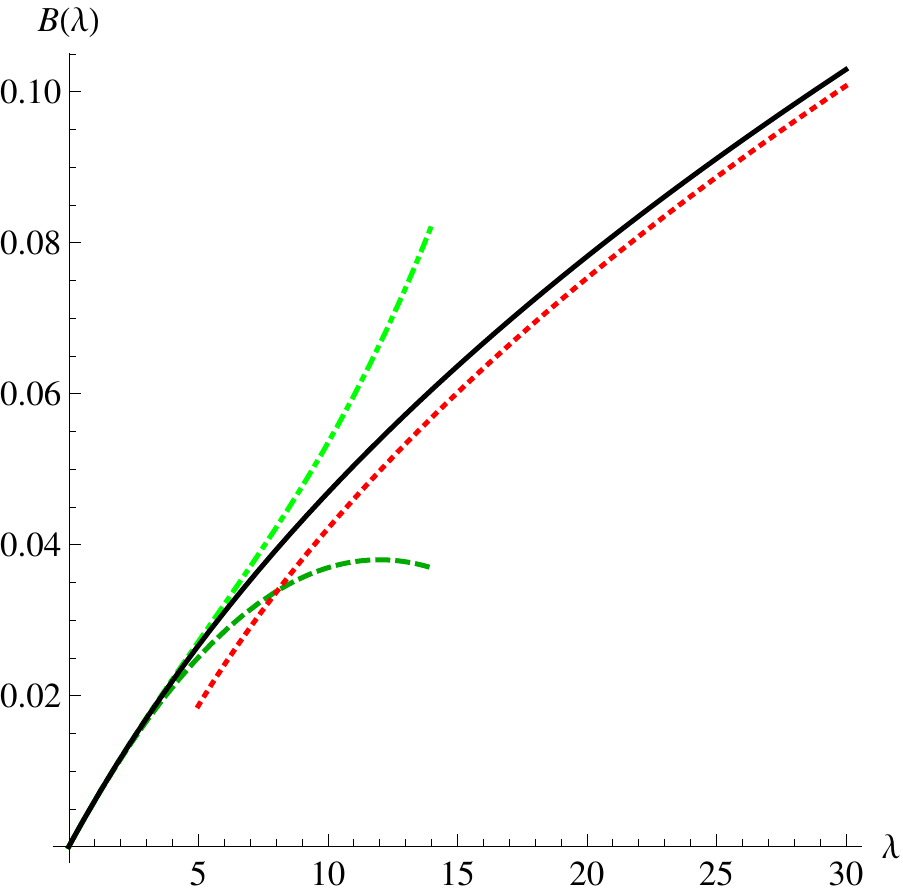}
\caption{Bremsstrahlung function $B(\lambda)$ (solid black line) given in eq. (\ref{eq_bremsstrahlung}).
The dotted green lines on the left denote two-loop and three-loop approximations, respectively,
and the dotted red line on the right corresponds to the first two terms of the strong coupling approximation.}
\label{fig:bremsstrahlung}
\end{figure}

It turns out that there is a special limit of $\Gamma_{\rm cusp}$,
namely the small angle limit $|\theta^2 - \phi^2| \ll 1$,
which can be computed exactly in $\phi$, $\lambda$ and $N$.
Here we simply quote the result and refer the interested reader 
to \cite{Correa:2012at} for more details.
Specifying to the planar case $N\gg 1$, we have
\begin{eqnarray}\label{exactsmallangle}
\Gamma_{\rm cusp} &=& (\phi^2-\theta^2) \frac{1}{1-\phi^2/\pi^2} B(\tilde\lambda)  + \ldots \,,  %+ \cO( (\phi^2 - \theta^2)^2 ) \,, 
\end{eqnarray}
with
$\tilde{\lambda} = \lambda (1-\phi^2/\pi^2 )$, and
\begin{eqnarray}\label{eq_bremsstrahlung}
B(\lambda) &=& \frac{1}{4 \pi^2} \frac{ \sqrt{\lambda} I_{2}(\sqrt\lambda )}{I_{1}(\sqrt{\lambda})} +\cO(1/N^2) \,,
\end{eqnarray}
where $I_{j}$ is the modified Bessel function.

Expanding this exact result  to the third order in $\lambda$, one reproduces the coefficient
of $\xi$ in eqs. (\ref{result1}), (\ref{result2}) and (\ref{result3}), respectively.

One can also expand our exact result at strong coupling. This can be compared against taking
the limit $|\phi - \theta| \ll 1$ of the formulas in \cite{Drukker:2011za}. In both cases,
we find
\begin{eqnarray}
\Gamma_{\rm cusp} &=& (\theta-\phi) \, \frac{\sqrt{\lambda}}{2 \pi} \frac{\phi}{ \sqrt{\pi^2-\phi^2}} +\ldots \,.%+ \cO( (\theta-\phi)^2)\,.
\end{eqnarray}

For a constant coupling of the Wilson loop to scalars, i.e. $\theta=0$,
eq. (\ref{exactsmallangle}) becomes
\begin{eqnarray}
\Gamma_{\rm cusp} &=& - \phi^2 \, B(\lambda) + \cO(\phi^4) \,,\quad  \theta = 0\,.
\end{eqnarray}
This is related to the radiation emitted by a quark undergoing a small change in direction \cite{Correa:2012at}.
The ``Bremsstrahlung'' function $B(\lambda)$ is plotted in Fig.~\ref{fig:bremsstrahlung}.

\section{SCALING LIMIT}
\label{sec:scaling}

The existence of the parameter $\theta$ allows us to define a new scaling limit.
Recall that $\Gamma_{\rm cusp}$ depends on $\cos \theta$ in a polynomial way.
It is easy to see  that the terms with the highest power $\cos^L \theta$ at $L$ loops
come from a diagram with $L$ scalar propagators ending on each Wilson
line. We can isolate such terms by taking the limit \cite{Correa:2012nk}
\begin{eqnarray}\label{scaling-limit}
\lambda \to 0\,,\quad e^{i \theta} \to \infty\,,\quad   {\kappa}=\frac{1}{4 \pi^2} \lambda e^{i \theta}\;\; {\rm fixed} \,.
\end{eqnarray}
In this way, we get a non-trivial function of $\phi$, $\kappa$ and $N$.
We see that at leading order (LO) in the limit, only ladder diagrams remain.
We will denote the leading order approximation by $\Gamma_{\rm cusp}^{\rm LO}$.

The sum of the ladder integrals satisfies a Bethe-Salpeter equation. Since we are only
interested in the leading UV divergence (of the sum of the ladder integrals), we can simplify the equation
to a one-dimensional Schr\"{o}dinger problem,
\begin{eqnarray}\label{eq_schroedinger}
\left[ - \partial_{y}^2 - \frac{\kappa}{8 } \frac{1}{\cosh y + \cos \phi } \right] \Psi(y) = -  \frac{\Omega^2}{4} \Psi(y) \,.
\end{eqnarray}
Here $y$ is a variable in an auxiliary space (related to relative positions on the two edges of the Wilson lines), and 
$\Gamma^{\rm LO}_{\rm cusp} = - \Omega_{0}$, where $\Omega_{0}$ is computed for the ground state of the system.

In the case of zero angle, the potential in eq. (\ref{eq_schroedinger}) becomes the integrable P\"{o}schl-Teller potential
$\cosh^{-2}(y/2)$, and one can solve for $\Gamma^{\rm LO}_{\rm cusp}$ in closed form.
One finds \cite{Correa:2012nk}
\begin{eqnarray}
\Gamma^{\rm LO}_{\rm cusp} &=& \frac{1- \sqrt{1+\kappa}}{2} \\
&&- \frac{\phi^2}{16} \kappa \left( \frac{1+\sqrt{1+\kappa}}{1+\kappa+2 \sqrt{1+\kappa}}\right) + \cO(\phi^4)\,. \nonumber
\end{eqnarray}

Moreover, for general angle, the problem can be solved perturbatively. 
One can show that to any order $L$, the LO answer for $\Gamma_{\rm cusp}$ can be written in terms
of harmonic polylogarithms of degree ${(2L-1)}$, and an algorithm to compute
the solution was given \cite{Henn:2012qz}. Moreover, the result was given explicitly up to six loops.
This calculation also confirmed analytically the $\xi^3$ term at three loops that was given 
in equation (\ref{result3}). 

There are a number of interesting properties of these results \cite{Henn:2012qz}.
A study of the explicit results to six loops shows that in fact only a 
subset of harmonic polylogarithms is needed.
We already mentioned in section \ref{sec:results} the absence 
of relative signs. Finally, when studying the limit $x\to0$ discussed
in section \ref{sec:limit}, one finds, at least up to six loops, that the asymptotic expansion
of $\Gamma_{\rm cusp}^{\rm LO}$ does not require multiple zeta values \cite{Blumlein:2009cf}
of depth two or higher.

One can also compare the solution of the Schr\"{o}dinger equation to
what one obtains from taking the scaling limit of the corresponding 
string theory result. A priori, the answer obtained does not have
to be the same, since the order of limits could be important. However,
the explicit calculation showed agreement \cite{Correa:2012nk}.

\begin{figure}
\centering\includegraphics[scale=0.75]{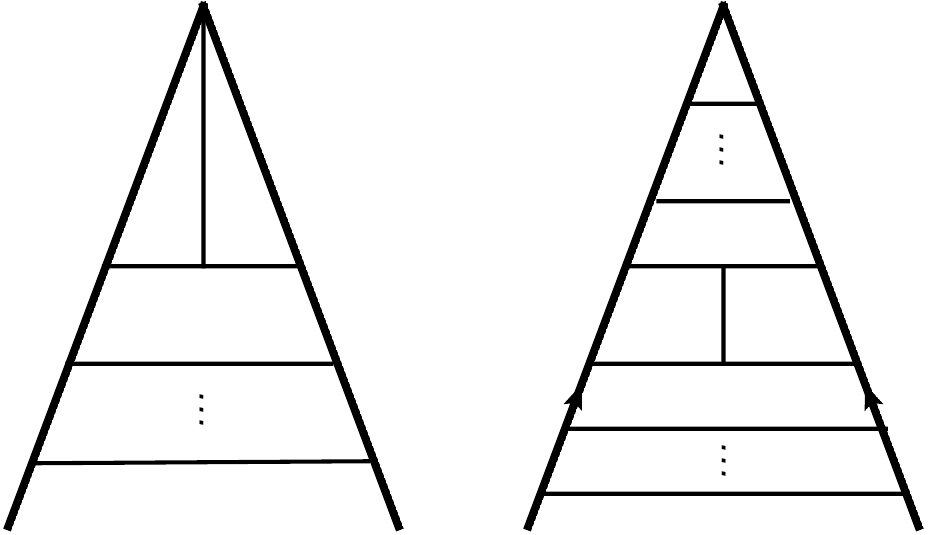}
\vspace{-0.3cm}
\caption{The two infinite classes of scalar integrals the contribute to $\Gamma_{\rm cusp}$ at NLO in the scaling limit (\ref{scaling-limit}).
The integral on the right contains a numerator factor $(k_1 + k_2)^2$, where $k_1$ and $k_2$ are the momenta flowing along
the propagators with the arrows, respectively.}
\label{fig:scaling}
\end{figure}

This analysis has recently been extended to the next-to-leading order (NLO)
in the scaling limit \cite{Henn:2012qz}. (See also
ref. \cite{Bykov:2012sc} for an analysis of the special case $\phi \to \pi$.)
One finds that at NLO there are two infinite classes of scalar integrals,
shown in Fig.~\ref{fig:scaling}. They satisfy modified Bethe-Salpeter
equations and can be computed algorithmically \cite{Henn:2012qz}.

\section{DISCUSSION} 
\label{sec:conclusion}

We have reviewed various new results for
the cusp anomalous dimension in $\mathcal{N}=4$ super Yang-Mills.

Recently, it was shown how to apply integrability techniques to
this problem \cite{Correa:2012hh,Drukker:2012de,Gromov:2012eu}. 
It would be interesting to reproduce the results
obtained here using these methods.

While the results presented here are valid for supersymmetric gauge 
theories, the structures found are also be relevant for QCD. 
Interesting extensions of our work would be the calculation
of the three-loop value of the cusp anomalous dimension in QCD,
as well as the first non-planar correction, which appears for the
first time at four loops.

\section*{ACKNOWLEDGMENTS}
This talk is based on the publications \cite{Correa:2012at,Correa:2012nk}
with D.~Correa, J.~Maldacena and A.~Sever, and on \cite{Henn:2012qz}
with T.~Huber.
We wish to thank the organizers of the conferences where this talk
was presented for their invitation, and the participants for many 
interesting discussions and comments.
JMH was supported in part by the Department of Energy grant DE-FG02-90ER40542.

\end{document}